\documentclass[final,5p,times,twocolumn,number,compress]{elsarticle}

\usepackage{amssymb}
\usepackage{amsmath}
\PassOptionsToPackage{hyphens}{url}
\usepackage[usenames,dvipsnames]{color}
\usepackage{setspace}
\usepackage{caption}
\usepackage{subcaption}
\usepackage{ulem}
\usepackage[T1]{fontenc}

\usepackage{hyperref}
\hypersetup{
    citecolor=blue,
    colorlinks=true,
    linkcolor=blue,
    filecolor=blue,      
    urlcolor=black,
    pdfpagemode=FullScreen,
}
\usepackage{siunitx}
\usepackage{physics}

\newcommand{\gev}{\giga\electronvolt}

\begin{document}

\title{Branching ratios for Higgs-mixed scalars at the GeV scale from hadronisation models with conservation laws}

\tnotetext[t1]{This article is registered under preprint number: KA-TP-33-2025, TTP25-037, P3H-25-078}

\author{Stefan Gieseke$^a$}
\ead{stefan.gieseke@kit.edu}

\author{Felix Kahlhoefer$^{b,c}$}
\ead{kahlhoefer@kit.edu}

\author{Henry Seebach$^b$}
\ead{henry.seebach@mail.de}

\address[1]{Institute for Theoretical Physics (ITP), Karlsruhe Institute of Technology (KIT), \\D-76131 Karlsruhe, Germany}
\address[2]{Institute for Theoretical Particle Physics (TTP), Karlsruhe Institute of Technology (KIT), \\D-76131 Karlsruhe, Germany}
\address[3]{Institute for Astroparticle Physics (IAP), Karlsruhe Institute of Technology (KIT), Hermann-von-\\Helmholtz-Platz 1, 76344 Eggenstein-Leopoldshafen, Germany}

\begin{abstract}

We investigate the decay modes of a CP-even scalar boson $\phi$ that mixes with the Standard Model Higgs boson, focusing on the mass range between 2 GeV and $2 m_\tau$. Starting from a higher-order perturbative calculation of the inclusive decays $\phi \to gg$ and $\phi \to s\bar{s}$, we employ a hadronisation model to obtain predictions for individual hadronic final states. Our hadronisation model is based on the Herwig cluster model, but incorporates various conservation laws to determine the allowed final states and their respective weights. The model includes two tunable parameters, which we determine using dispersion relation results at $m_\phi = 2$ GeV, enabling extrapolation to higher masses. Our predictions show that two-particle hadronic final states like $\pi^+ \pi^-$ and $K^+ K^-$ dominate over $\mu^+ \mu^-$ for $m_\phi$ near 2 GeV, suggesting promising targets for future experimental searches.

\end{abstract}

\maketitle

\section{Introduction}

Calculating the decay modes of a CP-even scalar particle in the GeV mass range has challenged physicists for decades, going back to the 1980s~\cite{Voloshin:1985tc} when it was still plausible that the Standard Model (SM) Higgs boson would have a mass below 10 GeV~\cite{gunionHiggsHuntersGuide2000}. While the problem is no longer relevant for the SM Higgs boson, it has received renewed interest in the context of new scalar particles that mix with the SM Higgs boson, obtaining the same coupling structure suppressed by a mixing angle $\theta$~\cite{pospelovSecludedWIMPDark2008,Batell:2009di}. Such particles are of great interest in the context of dark matter physics~\cite{Krnjaic:2015mbs,Bondarenko:2019vrb} and provide exciting targets for accelerator experiments~\cite{Beacham:2019nyx,Antel:2023hkf}. While decays into leptons can easily be calculated perturbatively, more complicated methods are needed to determine the decays into hadrons~\cite{bezrukovUncertaintiesHadronicScalar2018}. This problem is much more severe for CP-even scalars than for pseudoscalars, which are expected to mix with well-known QCD resonances~\cite{Ovchynnikov:2025gpx}, and for vectors, which can be studied experimentally using off-shell photons~\cite{Ilten:2018crw}. None of these possibilities exist for CP-even scalars, because the SM contains neither a light fundamental scalar nor narrow scalar QCD resonances at the GeV scale.

Over the years, various methods have been developed to address these issues. One possible is to study partial decay widths, such as $\phi \to \pi \pi$ or $\phi \to K K$ using dispersive methods, i.e.\ the analysis of scattering data using the optical theorem~\cite{Raby:1988qf}. The other possibility are higher-order perturbative calculations of inclusive decays such as $\phi \to g g$ or $\phi \to s \bar{s}$~\cite{winklerDecayDetectionLight2019}. Unfortunately, neither of these methods is complete: The dispersive method cannot give the total decay width, which would need to include additional final states such as $\phi \to \eta \eta$, while the perturbative method cannot give partial decay widths into exclusive final states. Even more unfortunately, the two methods cover different mass regions: Given the available experimental data, the dispersive method is reliable only up to approximately $2 \, \mathrm{GeV}$~\cite{blackstoneHadronicDecaysHiggsmixed2024}, which corresponds approximately to the lower bound up to which a perturbative expansion is expected to converge. As a result, there is currently no method that can predict the branching ratios of a CP-even scalar in the GeV mass range.

A possible solution could be to combine a perturbative calculation above $2 \, \mathrm{GeV}$ with a hadronisation model as implemented in parton shower generators such as \textsc{Herwig}~\cite{bewickHerwig73Release2024} or \textsc{PYTHIA}~\cite{bierlichComprehensiveGuidePhysics2022}. Such a hadronisation model would translate the ``hard'' process (such as a decay into a gluon pair) into hadronic final states. However, none of the existing tools are actually fit for this purpose. The reason is that conservation laws, which are decisive for understanding the decay patterns of particles at the GeV scale, are not explicitly enforced in the code. While this is not a problem in the energy range these tools are intended for, it leads to unphysical predictions for the scenario that we are interested in.

In the present work we address this shortcoming by proposing a new approach to calculate branching ratios for CP-even scalars: A hadronisation model that explicitly checks and enforces various conservation laws that are expected to hold for strong interactions. Combined with a perturbative calculation of the total decay width, this approach enables us to predict branching ratios for exclusive final states. Of course, the hadronisation model comes with unknown parameters that need to be determined from data. Fortunately, such data exists in the form of the partial decay widths obtained from the dispersive method. Requiring that the two methods should agree for $m_\phi = 2 \, \mathrm{GeV}$, we can determine the free parameters and obtain new predictions for the scalar decay modes in the mass range $2 \, \mathrm{GeV} < m_\phi < 2 m_\tau$, where the upper bound corresponds to the scalar mass where decays into tau leptons and $D$ mesons start to dominate the total width.

Our results have immediate implications for experimental searches for light CP-even scalars. The Belle II-experiment, for example, has recently performed a search for scalar resonances produced in the decay $B \to K \phi$. While the search has been carried out in a multitude of different final states, the lack of theoretical predictions for the branching ratios meant that only the di-muon final state could be used to constrain specific models, such as light scalars with Higgs mixing. The branching ratios that we calculate enable us to reinterpret the model-independent results from Belle II for various final states, finding that in particular the $\phi \to KK$ decay offers a promising target for future searches.

The remainder of this work is structured as follows. We briefly review the existing methods for calculating partial and total decay widths of a CP-even scalar with Higgs mixing in section~\ref{sec:review}. In  section~\ref{sec:hadron} we then discuss the available parton shower tools and their shortcomings, before introducing our own approach. We also discuss how we use existing results to determine the free parameters of our model. Our results are presented in section~\ref{sec:results}.

\section{Review of existing techniques}
\label{sec:review}

The model of a Higgs-mixed scalar (also called Higgs portal) is an extension of the SM with one additional real scalar field $S$. Due to interaction terms in the scalar potential, $S$ mixes with the SM Higgs boson $h_0$ after electroweak symmetry breaking. The physical mass eigenstates $h$ and $\phi$ are then obtained by an orthogonal rotation
\begin{equation}
    \begin{pmatrix} h_0 \\ S \end{pmatrix} = \begin{pmatrix} \cos \theta & \sin \theta \\ -\sin \theta & \cos \theta \end{pmatrix} \begin{pmatrix} h \\ \phi \end{pmatrix} \; .
\end{equation}
As a result, the usual interaction terms of the SM Higgs boson are suppressed by a factor of $\cos \theta$, while the new scalar $\phi$ obtains the same interaction terms with a factor of $ s_\theta \equiv \sin \theta$. As a result, $\phi$ couples to all massive particles of the SM proportionally to their mass. Since measurements of the production and decay modes of $h$ agree well with SM predictions, we know that $s_\theta \ll 1$, which makes it possible for $\phi$ to be much lighter than $h$ and still evade experimental detection. 

In the absence of additional decay modes, the partial and total decay widths are simply given by those of a SM Higgs boson with mass $m_\phi$, multiplied by $s_\theta^2$. We can therefore use many of the standard results for Higgs boson decays from the literature. However, these results usually rely on approximations based on the Higgs boson being light compared the top quark and very heavy compared to all other quark flavours, which need to be reassessed in the context of a light scalar. 

\subsection{Perturbative decays} 
\label{sec:perturbative}

The leading order decay of $\phi$ into a pair of gluons proceeds via a quark triangle diagram. The corresponding decay width is given by \cite{spiraHIGGSBOSONPRODUCTION1995} 
\begin{equation} \label{eq:gg_LO}
    \Gamma_{\mathrm{LO}}^{gg} = \frac{G_F \alpha_s^2 s_\theta^2}{36\sqrt{2}\pi^3} m_\phi^3 \left|\frac{3}{4}\sum_q A_q(\tau_q)\right|^2,
\end{equation}
where the sum extends over all quarks running in the loop and $A_q$ is given by
\begin{equation}
    A_q(\tau_q) = 2\frac{\tau_q + (\tau_q-1)f(\tau_q)}{\tau_q^2},
\end{equation}
with 
\begin{equation}
    f(\tau)=\begin{cases}
        \arcsin^2 \sqrt{\tau} & \tau \leq 1 \\
        -\frac{1}{4} \left(\log{\frac{1+\sqrt{1-\tau^{-1}}}{1-\sqrt{1-\tau^{-1}}}} - i\pi \right)^2 & \tau > 1
    \end{cases}
\end{equation}
and $\tau_q = m_\phi^2 / (4m_q^2)$. Since $\lim_{\tau\to\infty}A(\tau)=0$, light flavours can be excluded from the sum over $q$. For scalar masses above 2 GeV, these are up, down and strange quarks.

The decay width of the Higgs boson to gluons is currently known up to $\text{N}^4\text{LO}$ \cite{herzogHiggsDecaysHadrons2017} in the limit of infinite top quark mass and five massless flavours. This approximation is well-justified for the SM Higgs boson, but is questionable for scalars at the GeV scale, since charm and bottom mass effects could be sizeable. Only the NLO correction is known including the full mass dependence of the three heavy flavours \cite{spiraHIGGSBOSONPRODUCTION1995} and is given by
\begin{equation} \label{eq:gg_NLO}
    \Gamma_{\mathrm{NLO}}^{gg} = \Gamma_\mathrm{LO}^{gg} \left(1+E \frac{\alpha_s}{\pi} \right)
\end{equation}
with
\begin{equation} \label{eq:NLO_fac}
    E = \frac{95}{4} - \frac{7}{6} n_f + \frac{33-2n_f}{6} \log \frac{\mu^2}{m_\phi^2} + \Delta E,
\end{equation}
where $\Delta E$ involves some numerical integrals given in the appendix of \cite{spiraHIGGSBOSONPRODUCTION1995}. These can be calculated numerically with the public code \textsc{higlu} \cite{spiraHIGLUProgramCalculation1995}.

The parameter $n_f$ describes the number of light ("active") quark flavours and the on-shell renormalization scheme is used for the quark masses. \textsc{Higlu} takes as quark mass inputs the $\overline{\text{MS}}$ masses $m_c(\mu=\SI{3}{\gev})=\SI{0.98}{\gev}$ and $m_b(m_b)=\SI{4.18}{\gev}$. These values are taken from Ref.~\cite{navasReviewParticlePhysics2024}, where the charm mass is given as $m_c(m_c)=\SI{1.27}{\gev}$, which is then evolved to a scale of $\SI{3}{\gev}$ using \textsc{rundec-python}\footnote{This code is a python wrapper of the C++ program \textsc{rundec} \cite{schmidtCRunDecPackageRunning2012} and can be found at \url{https://github.com/DavidMStraub/rundec-python}.}. We also use this package for the running of the strong coupling constant. The on-shell masses are calculated by \textsc{higlu} to be $m_c^\text{OS}=\SI{1.43}{\gev}$ and $m_b^\text{OS}=\SI{4.83}{\gev}$. The top mass in the on-shell scheme is taken to be $m_t=\SI{172.5}{\gev}$ \cite{navasReviewParticlePhysics2024}. 

The number of active quark flavours $n_f$ appears explicitly in the decay width through eq.~\eqref{eq:NLO_fac}, but is also implicitly contained in $\alpha_s^{(n_f)}(\mu)$ because it affects the running of the coupling. 
For the gluonic decay width, $n_f$ is related to the number of quarks contributing to the real corrections $\phi\to gg\to g q\bar{q}$, where one gluon splits into a quark-antiquark pair. Here we set $n_f=3$, which effectively subtracts the $\phi\to g c\bar{c}$ and $\phi \to g b\bar{b}$ contributions from the decay width but leaves behind logarithms of the form $\log(\mu^2/m_c^2)$ and $\log(\mu^2/m_b^2)$. It is argued in Ref.~\cite{spiraQCDEffectsHiggs1997} that these can be resummed by going from $\alpha_s^{(5)}$ to $\alpha_s^{(3)}$, thereby decoupling the charm and bottom quarks from the theory. Although the resummation is not necessary for the convergence of the perturbation series, because the logarithms are not large in the mass range that we consider, we include it for consistency. 

Higher order corrections to $\phi\to gg$ exist only in the approximation of infinite top mass and zero charm and bottom mass. However, already at NLO the impact of the charm and bottom quarks are found to be very small, with the exact $K$ factor $K_\text{NLO}^\text{exact}=\Gamma_\text{NLO}^{gg}/\Gamma_\text{LO}^{gg}$ differing from the approximate one including only top quarks by at most 6\%. We take this as justification for including  higher-order contributions up to $\text{N}^4\text{LO}$ in the limit of heavy top quark and massless other flavours in the decay width, while keeping the exact quark mass dependence up to NLO.

\begin{figure}[t]
    \centering
    \includegraphics[width=0.8\columnwidth]{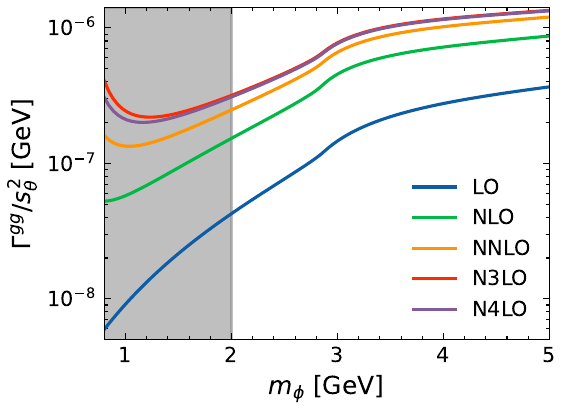}
    \hfill
    \includegraphics[width=0.8\columnwidth]{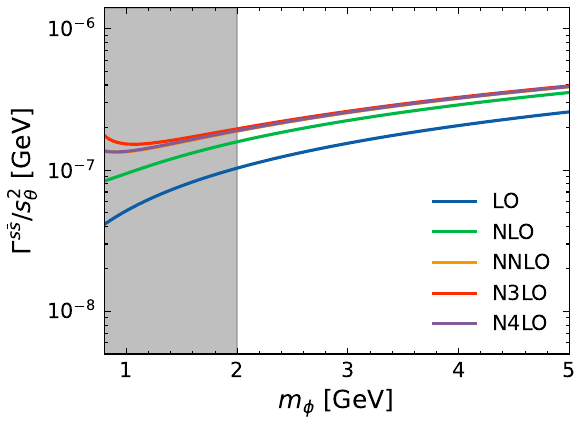}
    \caption{Decay width into gluons $\Gamma^{gg}$ (top) and into strange quarks $\Gamma^{s\bar{s}}$ (bottom) as a function of the scalar mass $m_\phi$ at different orders in perturbation theory.}
    \label{fig:all_orders}
\end{figure}

The decay width as a function of the scalar mass is shown in the top panel of figure \ref{fig:all_orders} for different orders in perturbation theory. As expected, higher order corrections become consecutively smaller, indicating perturbative convergence above $m_\phi=\SI{2}{\gev}$. Below $\SI{2}{\gev}$, on the other hand, the higher order corrections start to move apart again and the decay width diverges around $m_\phi=\SI{1}{\gev}$, due to the divergence of $\alpha_s$. We emphasize that the gluonic decay width at N$^4$LO is a factor of 3--5 larger than the leading-order estimate, which is commonly used by the community~\cite{ferberDarkHiggsBosons2024}. This has important implications for experimental searches that target the decay $\phi \to \mu^+ \mu^-$, as performed for example by LHCb~\cite{LHCb:2016awg} or CMS~\cite{CMS:2021sch}. Using the leading-order estimate of the gluonic decay width strongly overestimates the branching ratio into muons, and hence the strength of experimental constraints.

In the results above, we have set the renormalization scale to $\mu=m_\phi$. The dependence of the decay width on $\mu$ can be used as a measure of missing higher-order contributions, because the exact result should have no $\mu$ dependence. To estimate the theoretical error of our calculation, we therefore compare the result for $\mu = m_\phi$ to the one obtained for  $\mu=2m_\phi$.\footnote{A more common approach would be to vary $\mu$ between $\tfrac{1}{2}m_\phi$ and $2 m_\phi$. However, doing so overestimates the uncertainty, since $\alpha_s$ diverges around 1 GeV. We have checked that randomly varying the renormalisation scale between $m_\phi$ and $2 m_\phi$ gives a similar uncertainty estimate as the one obtained from our approach.} For $m_\phi = 2 \, \mathrm{GeV}$ and the N$^4$LO result, we obtain a relative uncertainty of 24\%. This uncertainty will be included in our calculations below. 

For our analysis, we also need the decay width for $\phi \to s \bar{s}$ which plays a relevant role in the production of kaons. At leading order, this decay width is given by \cite{gunionHiggsHuntersGuide2000}
\begin{equation} \label{eq:qq_lo}
    \Gamma_\text{LO}^{s\bar{s}}=\frac{3s_\theta^2 G_F m_\phi m_s^2}{4\sqrt{2}\pi}\beta^3
\end{equation}
with $\beta=\sqrt{1-4m_s^2/m_\phi^2}$.\footnote{Ref.~\cite{winklerDecayDetectionLight2019} suggests to replace $m_s$ by $m_K$ in the expression for $\beta$ in order to correctly capture the closure of the phase space for $m_\phi \to 2 m_K$. Here we stick to a purely perturbative calculation, noting that the result becomes unphysical for $m_\phi < 2 m_K$.}
In Ref.~\cite{herzogHiggsDecaysHadrons2017} the QCD corrections to this process have been calculated up to $\text{N}^4\text{LO}$ in the limit of vanishing masses of the light quarks and infinite top mass, which is a reasonable approximation for decays into strange quarks. A comparison of the decay width at different orders of perturbation theory is shown in the bottom panel of figure \ref{fig:all_orders}. While the first two corrections are large, the higher orders provide only small corrections, which is a sign of perturbative convergence above 2 GeV. 

\subsection{Dispersion relations}
\label{sec:dispersion}

Since QCD perturbation theory only gives reliable results for scalar masses above approximately $\SI{2}{\gev}$, other tools are needed to calculate hadronic decay widths for smaller scalar masses. Below the chiral symmetry breaking scale of about $\SI{1}{\gev}$, chiral perturbation theory can be used to directly calculate the decays into pions and kaons~\cite{winklerDecayDetectionLight2019}. In the intermediate region between, the decay widths can be obtained from dispersion relations~\cite{bezrukovUncertaintiesHadronicScalar2018}. The most recent calculation was made in Ref.~\cite{blackstoneHadronicDecaysHiggsmixed2024}, which calculates the decay to pions and kaons in a two channel approximation. 

\begin{figure}[t]
    \centering
    \includegraphics[width=\columnwidth]{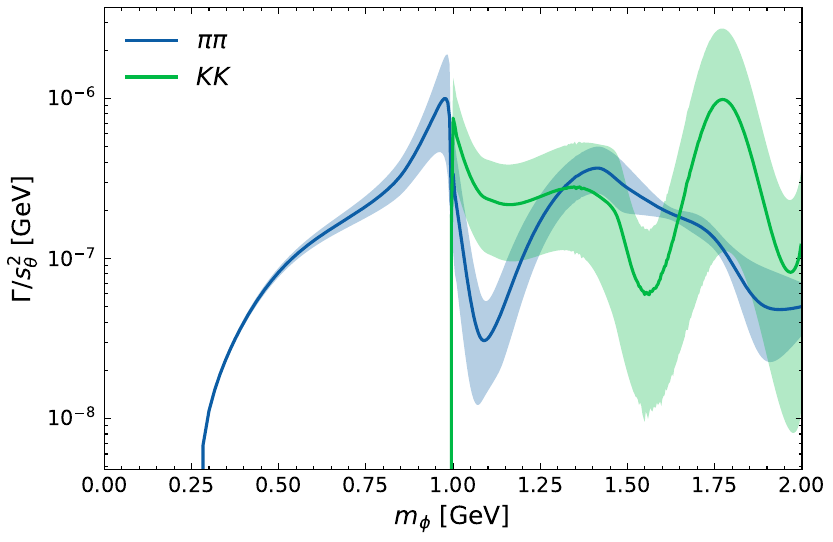}
    \caption{Widths for the decay of $\phi$ to pions and kaons obtained from dispersion relations in Ref.~\cite{blackstoneHadronicDecaysHiggsmixed2024}.}
    \label{fig:dispersive}
\end{figure} 

The resulting decay widths to pions and kaons are shown in figure \ref{fig:dispersive}. The plot is generated using the code \textsc{hipsofcobra}, published together with Ref.~\cite{blackstoneHadronicDecaysHiggsmixed2024}, using the conditions proposed in Ref.~\cite{Donoghue:1990xh} to match to chiral perturbation theory at low energies. 
The decay widths show several pronounced peaks due to scalar resonances, which enter through the experimental data of $\pi\pi\to\pi\pi$ and $\pi\pi\to KK$ scattering. The uncertainty bands come from an estimation of the uncertainties of the chiral perturbation theory results used for the matching, as well as from the errors in the experimental data. Since the available data only extends up to a centre-of-mass energy of $\sqrt{s}=\SI{2}{\gev}$, this approach can only be reliably used for $m_\phi \leq 2 \, \mathrm{GeV}$.

The calculation in Ref.~\cite{blackstoneHadronicDecaysHiggsmixed2024} is performed entirely in the two-channel approximation, neglecting all other decay channels of $\phi$ that might be important, such as e.g. $\phi \to \eta\eta$. It is argued in Ref.~\cite{bezrukovUncertaintiesHadronicScalar2018} that this reduction to a two-channel system may introduce significant uncertainties. It should be kept in mind that no estimate of the size of these errors exists so far, so the uncertainty associated with the pion and kaon decay widths may be larger than what is shown in figure \ref{fig:dispersive}.
Nevertheless, we will use these decay widths and their quoted uncertainties to fix the open parameters of the hadronisation model that will be introduced in the next section.

\section{Hadronisation model}
\label{sec:hadron}

To obtain the partial decay widths into pions and kaons also for $m_\phi > 2 \, \mathrm{GeV}$, one could interface the perturbative calculation of $\phi \to gg$ and $\phi \to s\bar{s}$ presented above with the hadronisation models provided by parton shower generators such as \textsc{Herwig}~\cite{bewickHerwig73Release2024} or \textsc{PYTHIA}~\cite{bierlichComprehensiveGuidePhysics2022}. While \textsc{Herwig} is based on the cluster model for hadronisation~\cite{webberQCDModelJet1984}, \textsc{Pythia} uses the Lund string model \cite{anderssonPartonFragmentationString1983,sjostrandJetFragmentationMultiparton1984}.

It turns out, however, that neither of the two available methods give satisfactory results for such a low-energy system. The cluster model in \textsc{Herwig} assumes that each gluon decays non-perturbatively into a pair quarks, each of which ends up in a separate meson, such that it is impossible to obtain two-meson final states. Indeed, for $m_\phi = 2\,\mathrm{GeV}$ the most common final state in the \textsc{Herwig} simulation is $\pi^+ \pi^- 2\pi^0$, accounting for approximately 12\% of all events. In \textsc{Pythia}, on the other hand, two particle final states are possible, but the most common final state turns out to be $\pi^+ \pi^- \pi^0$ (25\% of all events). This final state, however, is CP-odd and should never arise in strong decays of a CP-even scalar. This conservation law is however  ignored in \textsc{Pythia}, such that the output contains various unphysical final states. 

To address these shortcomings, we have developed a new hadronisation model, based on the cluster model in \textsc{Herwig}. For the mass range that we are interested in, we assume that no parton shower takes place and that the decay products (either gluons or strange quarks) immediately form a single cluster.  This is in contrast to the original approach, where gluons decay non-pertubatively into a quark-antiquark pair.  In a high-energy environment this allows to follow the colour structure of the event after the parton shower in order to form primary colour singlet clusters.  In our case the gluon pair from the decay is already in a colour-singlet state and hence it is conceivable that this would directly form a single cluster.  In the targeted mass range, this cluster would be a light cluster and not undergo any cluster fission.  The resulting cluster is then further
decayed into hadrons based on a number of simple assumptions: First, we consider only decays into mesons. In the mass range under consideration, decays into baryon-antibaryon pairs are expected to be suppressed, both because of the small available phase space and because of the need to create two quark-antiquark pairs from the plasma. Furthermore, the cluster is assumed to always decay into exactly \textit{two} mesons initially. Although decays into more particles are possible in principle, they are again expected to be suppressed due to the smaller available phase space.

The relative proportions in which different meson pairs are produced are determined solely by symmetry considerations. We start with all possible meson combinations\footnote{We exclude the $f_0(500)$ and $K_0^*(700)$ resonances, because their widths and decay modes are not sufficiently well known. Other scalar resonances, such as $f_0(980)$, are included and treated as quark-antiquark bound states consisting of up and down quarks only.} that conserve charge, flavour, parity, G-parity, charge conjugation symmetry, angular momentum and isospin as outlined in~\ref{app:selection_rules}. Each pair of mesons $m_1$ and $m_2$ is then assigned a weight given by
\begin{equation}
    \label{eq:weights}
W(m_1,m_2)=p(m_1,m_2,m_\phi) W_q W_v W_I W_\mathrm{sym}.
\end{equation}
To obtain the probability for $\phi$ to decay into a particular meson pair, each weight is normalized by dividing by the sum of all weights. The individual factors in eq.~\eqref{eq:weights} are 
\begin{itemize}
    \item $p(m_1,m_2,m_\phi)$: A phase-space factor. It corresponds to the momentum of the two mesons with masses $m_1$ and $m_2$ in the rest frame of $\phi$~\cite{navasReviewParticlePhysics2024}, which is given by
    \begin{align} \label{eq:momentum}
        p(m_1,m_2,m_\phi)=\tfrac{1}{2m_\phi} &\left[(m_\phi^2-(m_1+m_2)^2)\right. \nonumber \\ & \; \left.(m_\phi^2-(m_1+m_2)^2)\right]^{\frac{1}{2} }.
    \end{align}
    \item $W_q$: A weight assigned for the quark content of the mesons. It consists of a weight $w_q$ for each quark-antiquark pair of flavour $q$, which quantifies the probability of creating this pair from the vacuum. We will discuss this factor in more detail below.
    \item $W_v$: A weight counting the spin multiplicity of the final state, see \ref{sec:angularmom} for details. If the final state does not carry orbital angular momentum, conservation of angular momentum implies $W_v = (2 j_1 + 1)$, where $j_1$ denotes the spin of $m_1$, which must be equal to the spin of $m_2$. In our analysis, we also allow for final states with non-zero angular momentum, which are however suppressed by a factor $0 < a_v < 1$. As a result, we set
    $W_v= (1-a_v)(2j_1+1)+a_v(2j_1+1)^2$ if $j_1=j_2$ and $P_1=P_2$, where $P_{1,2}$ denotes the parity of the two mesons, and $W_v = a_v(2j_1+1)(2j_2+1)$ otherwise.
    \item $W_I$: This weight takes into account isospin conservation as explained in~\ref{sec:isospin}. The selection rules require that both mesons have the same isospin $I$. Decays are therefore suppressed by the factor $            W_I=1/(2I+1)$.
    \item $W_\mathrm{sym}$: A factor to take into account the multiplicity of the final state. We assign  each pair of non-identical mesons a factor of $W_\text{sym} = 2$ and each pair of identical mesons a factor of $W_\text{sym} = 1$.
\end{itemize}

\subsection{Weight calculation}

To calculate $W_q$ we assume that for the gluon channel, two quark-antiquark pairs of arbitrary flavour are created. For charged mesons $m_1$ with quark content $(q_1,\bar{q}_2)$ and $m_2$ with $(q_2,\bar{q_1})$, the quark weight is then
\begin{equation}
    W_q=w_{q_1}w_{q_2} \; .
\end{equation}
We assume isospin symmetry, such that $w_u = w_d$. Since the final weights will be normalised, we can set both weights equal to unity without loss of generality. The only free parameter that we need to introduce is therefore the strange-quark weight $w_s$.

For a pair of (flavour) neutral mesons, which may be in a superposition of quark-antiquark states,
\begin{equation}
    W_q=\sum_q p_q^{m_1}p_q^{m_2}w_q^2 \; ,
\end{equation}
where  the sum runs over all quarks flavours (up, down and strange) and $p_q^m$ is the probability of finding the quark-antiquark pair of flavour $q$ in meson $m$. For the light neutral mesons, we take these probabilities as~\cite{bahrHerwigPhysicsManual2008}:
\begin{align}
    p_{ud}^{\pi^0} & = 1 \, , &     p_{ud}^{\eta} & = \cos^2(\theta + \varphi) \, , & p_{ud}^{\eta'} & = \sin^2(\theta + \varphi) \, ,\\
    p_{s}^{\pi^0} & =0 \, ,  & p_{s}^{\eta}& =\sin^2(\theta + \varphi)  \, ,
    & p_{s}^{\eta'} & =\cos^2(\theta + \varphi)
\end{align}
with $\varphi=\arctan(\sqrt{2})$. The probabilities for up and down quarks are combined into one, such that $p_{ud}^{\pi^0}$ is the probability of finding $u\bar{u}$ or $d\bar{d}$ in a $\pi^0$ meson. We implement this mixing for $(\pi^0,\eta,\eta')$ with $\theta=\SI{-23}{\degree}$ and for $(\rho^0,\Phi,\omega)$ with $\theta=\SI{36}{\degree}$ as in the source code of \textsc{Herwig 7.3} \cite{bewickHerwig73Release2024}.

In the strange-quark channel, only strange mesons can be produced, and it is assumed that one $s\bar{s}$ pair is already present. A weight is only assigned for the other quark pair. For charged mesons $m_1$ with quark content $(s,\bar{q})$ and $m_2$ with $(q,\bar{s})$ this means simply 
    \begin{equation}
        W_q=w_q
    \end{equation}
and for neutral mesons
    \begin{equation}
        W_q=p_s^{m_1}p_s^{m_2} w_s.
    \end{equation}

The procedure outlined above generates a list of weights for the allowed decays of $\phi$ for both the gluon and the strange-quark channel. The branching ratios from both channels are then added in proportion:
\begin{align}
    \text{BR}(\phi \to m_1,m_2)= & \frac{\Gamma(\phi\to gg)}{\Gamma_\mathrm{had}}W_{gg}(m_1,m_2) \nonumber \\ & +\frac{\Gamma(\phi\to s\bar{s})}{\Gamma_\mathrm{had}} W_{s\bar{s}}(m_1,m_2)
\end{align}
where $\Gamma_\mathrm{had}=\Gamma(\phi\to gg)+\Gamma(\phi\to s\bar{s})$ denotes the total hadronic decay width and the weights are assumed to be normalised. After generating all initial meson pairs with their associated branching ratios, each meson is further decayed until a sufficiently long-lived final state is reached. As we are interested in the hadronic decay width and branching ratios for particular final states, no explicit decay kinematics is needed.  Therefore all further computations can be done independent from a full implementation of this model into \textsc{Herwig}.  

\subsection{Fitting the model to data} \label{sec:fit}

The hadronisation model described above introduces two free parameters: the strange quark weight $w_s$ and the suppression of orbital angular momentum $a_v$. These two parameters can be determined by comparing the predictions of our model to the results from the dispersion relations discussed in section~\ref{sec:dispersion}. This comparison is done at $m_\phi = 2 \, \mathrm{GeV}$, where both methods are in principle applicable. Since the hadronisation model only predicts branching ratios, the comparison requires as additional input the total hadronic decay width. Given the sizeable theory uncertainties, as discussed in section~\ref{sec:perturbative}, we allow the hadronic decay width to deviate from the theory prediction by a factor $a_\Gamma$, which is constrained by the uncertainty determined from the scale dependence.

To determine the three free parameters, we construct a likelihood function. Since the errors $\sigma^\pm$ for $\Gamma_\pi$ and $\Gamma_K$ are very asymmetric, we follow the approach in Ref.~\cite{barlowAsymmetricErrors2024} and use the likelihood function 
\begin{equation}
    \log L(a;\hat{a},\sigma^\pm)=-\frac{1}{2} \left(\frac{a-\hat{a}}{\sigma + (a-\hat{a})\sigma'}\right)^2
\end{equation}
with $\hat{a}$ denoting the measured value and
\begin{equation}
    \sigma=\frac{2\sigma^+\sigma^-}{\sigma^++\sigma^-}, \qquad \sigma'=\frac{\sigma^+-\sigma^-}{\sigma^++\sigma^-} \; .
\end{equation}
This likelihood function generalises the Gaussian likelihood and satisfies the requirement $\log L(\hat{a})-\log L(\hat{a}\pm \sigma^\pm) = -\frac{1}{2}$.

The total log-likelihood is then given as the sum
\begin{equation}
    \begin{split}
    -2\log \mathcal{L} & (w_s,a_v,a_\Gamma) \\
    &= -2\log L\left(\Gamma_\pi(w_s,a_v,a_\Gamma);\widehat{\Gamma}_\pi,\sigma_\pi^\pm\right) \\
    &\hspace{4.5mm}- 2\log L\left(\Gamma_K(w_s,a_v,a_\Gamma);\widehat{\Gamma}_K,\sigma_K^\pm\right)  \\
    &\hspace{4.5mm}- 2\log L\left(\Gamma_\mathrm{had}(a_\Gamma);\widehat{\Gamma}_\mathrm{had},\sigma_\Gamma^\pm\right) \; .
    \end{split}
\end{equation}
The minimum of this log-likelihood determines the best-fit values of the model parameters. The corresponding uncertainties are obtained from the \textsc{MINOS} algorithm that is implemented in \textsc{iminuit}~\cite{jamesMinuitSystemFunction1975}, which performs a profile likelihood scan for all parameters and returns asymmetric upper and lower errors for each parameter. These uncertainties are then propagated to the output of our hadronisation model by varying each parameter within the quoted uncertainty interval and taking the largest and smallest prediction for each branching ratio.

We emphasize that even though the number of model parameters is equal to the number of constraints, it is not guaranteed that a good fit can be obtained. This is because both $a_v$ and $w_s$ are constrained to lie in the range $[0,1]$.

\section{Results}
\label{sec:results}

\begin{figure}[t]
    \centering
    \includegraphics[width=\columnwidth]{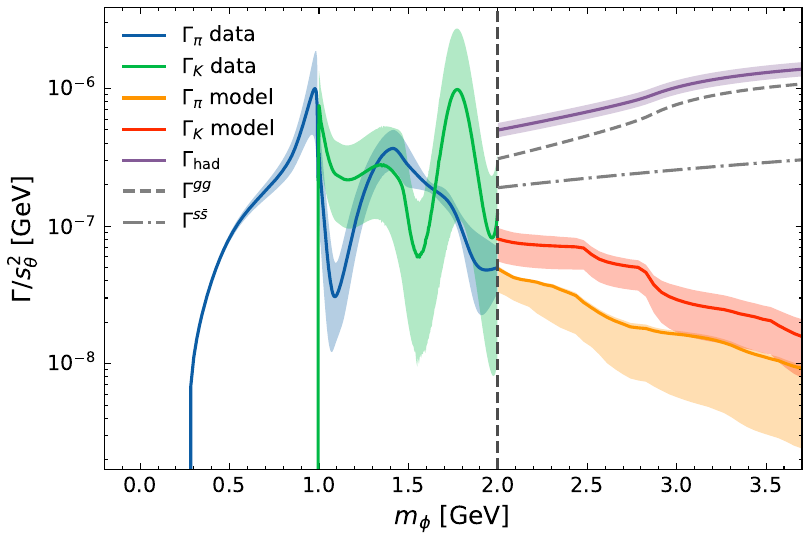}
    \caption{Decay widths of the CP-even scalar $\phi$ as a function of mass. Below 2 GeV, the pion and kaon decay widths are obtained from dispersion relations~\cite{blackstoneHadronicDecaysHiggsmixed2024}
(labelled as "data"). Above 2 GeV, the decay widths into gluons and strange quarks are obtained from a perturbative calculation, while the decay widths into pions and kaons are obtained from our hadronisation model, after fitting the free parameters to agree with the result from dispersion relations at 2 GeV.}
\label{fig:fit_result}
\end{figure} 

The fitting procedure described above gives the following results:
\begin{equation}
    \begin{aligned}
        w_s^0 &= 0.11^{+0.54}_{-0.11} \\
        a_v^0 &= 0.0^{+0.32} \\
        a_\Gamma^0 &= 1.0^{+0.12}_{-0.12} \\
        -2 \log \mathcal{L}^0 &= 0.13
    \end{aligned}
\end{equation}
Our likelihood function is normalised in such a way that $-2 \log \mathcal{L}^0 = 0$ would correspond to a perfect fit, while a $1\sigma$ deviation in a single observable would give $-2 \log \mathcal{L}^0 = 1$. The maximal value of the likelihood in our fit therefore indicates that the constraints can easily be satisfied within their uncertainties. 
This expectation is confirmed in figure~\ref{fig:fit_result}, which compares the partial widths into pions and kaons from dispersion relations for $m_\phi < 2 \, \mathrm{GeV}$ with the predictions of our best-fit hadronisation model for $m_\phi > 2 \, \mathrm{GeV}$. The two predictions are found to match very well at the boundary $m_\phi = 2 \, \mathrm{GeV}$.

We find that the best-fit strange-quark weight is only slightly larger than zero, corresponding to a substantial suppression for the creation of strange-quark pairs. This is because the prediction for $\Gamma_K$ from dispersion relations is quite small at $m_\phi = 2 \, \mathrm{GeV}$ compared to the perturbative prediction for the decay $\phi \to s \bar{s}$, such that the decay $\phi \to gg$ should not give a large contribution to $\Gamma_K$. Moreover, larger values of $w_s$ would suppress $\Gamma_\pi$. However, the results from dispersion relations have large uncertainties, and $\Gamma_K$ varies strongly as a function of $m_\phi$, such that larger values of $w_s$ may also be compatible with the available information. For comparison, the strange-quark weight used in \textsc{Herwig} is $w_s = 0.68$, which is approximately at the upper boundary of our confidence region for $w_s$.

Furthermore, our fit clearly prefers a strong suppression of orbital angular momentum, with the best-fit point lying at the boundary $a_v = 0$. This is because larger values of $a_v$ would reduce the partial width $\Gamma_\pi$ below the prediction from dispersion relations. For the same reason the uncertainty band for $\Gamma_\pi$ extends only to smaller values. In other words, within our hadronisation model, there is no freedom to increase $\Gamma_\pi$ beyond the value predicted by dispersion relations.

\begin{figure}[t]
    \centering
    \includegraphics[width=\columnwidth]{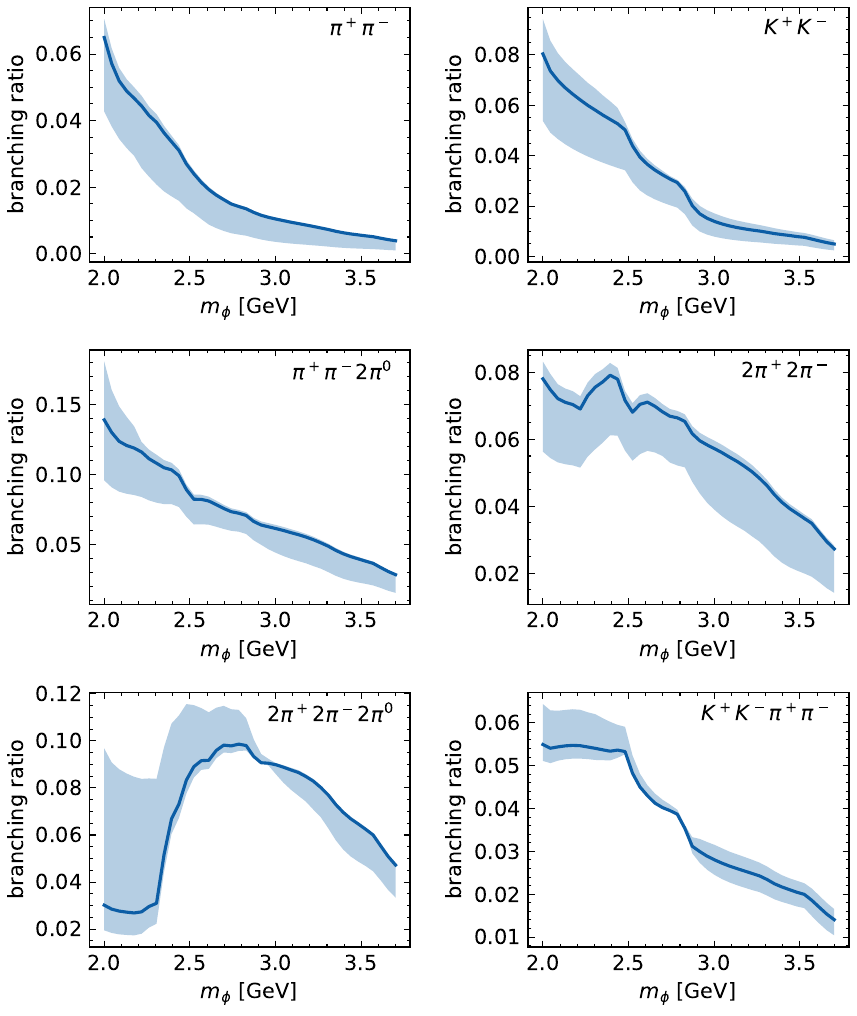}
    \caption{Branching ratios for the most relevant final states as a function of the scalar mass. In each case, the solid line corresponds to the best-fit prediction and the band represents the uncertainties resulting from the uncertainties in the partial widths obtained from dispersion relations at $m_\phi = 2 \, \mathrm{GeV}$.}
\label{fig:branchings}
\end{figure} 

As shown in figure~\ref{fig:fit_result}, our hadronisation model makes it possible to extrapolate the predictions for $\Gamma_\pi$ and $\Gamma_K$ from dispersion relations beyond scalar masses of 2 GeV. Moreover, we can also obtain predictions for other final states. These arise from two-body decays of the scalar into heavier mesons, which rapidly decay into lighter mesons, leading to higher-multiplicity final states. The predicted branching ratios for some of these final states are shown in figure~\ref{fig:branchings}. All of these final states contain at least two charged particles, allowing for the reconstruction of the decay vertex. Nevertheless, some of them also include neutral pions, which makes the reconstruction of the invariant mass of the decaying particle more challenging. Nevertheless, for all of the final states shown in figure~\ref{fig:branchings}, the branching ratios are larger than the one into muons (which varies between 3\% and 2\% in the mass range that we consider). As a result, these final states offer an attractive target for future searches.

To make this point more explicit, we derive sensitivity projections for Belle II based on the published search for $B^+ \to K^+ \phi$ followed by $\phi \to e^+ e^-$, $\mu^+ \mu^-$, $\pi^+ \pi^-$ or $K^+ K^-$. Belle~II has published both model-independent limits on $\text{BR}(B^+ \to K^+ \phi) \times \text{BR}(\phi \to X)$ as well as a reinterpretation of these limits in the context of a Higgs-mixed scalar. While the Belle II exclusion extends to scalar masses $m_\phi > 2 \, \mathrm{GeV}$, this result underestimates the hadronic decay width and hence overestimates the lifetime and the branching ratio into muons. With our new results, we find that the published Belle II search has no sensitivity to dark scalars above 2 GeV.\footnote{The same issue is expected to affect also the exclusion limit from LHCb~\cite{LHCb:2016awg}, which uses the same assumptions for the branching ratio into muons, as well as potentially the limit from CMS~\cite{CMS:2021sch}, which does not provide any details on the assumed branching ratios. However, the effect is expected to be milder than for Belle II, since the sensitivity of LHC-based experiments extends to smaller lifetimes as a result of the higher boost factors. Projections for the sensitivity of LHCb to Higgs-mixed scalars including hadronic final states have been derived in Ref.~\cite{Craik:2022riw}.}

\begin{figure}[t]
    \centering
    \includegraphics[width=0.9\columnwidth]{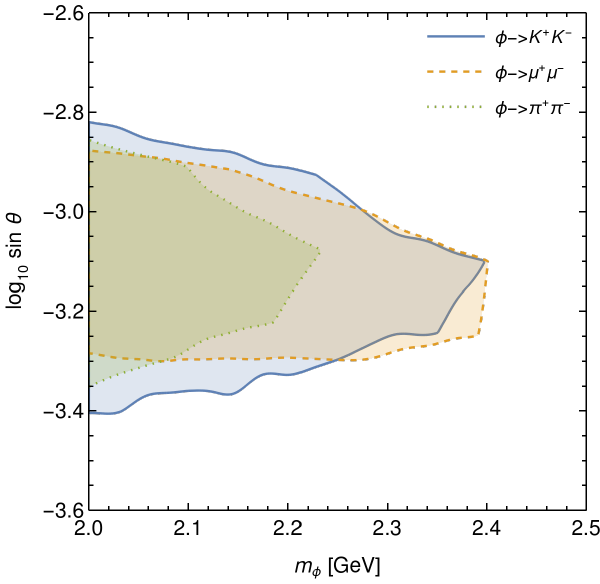}
    \caption{Sensitivity of Belle II to a Higgs-mixed scalar with mass $m_\phi$ and mixing angle $\sin \theta$ for different final states, assuming that the expected exclusion limits from Ref.~\cite{collaborationSearchLonglivedSpin02023} can be improved by a factor of 20.}
\label{fig:sensitivities}
\end{figure} 

Nevertheless, future searches based on more data may be able to probe into this difficult mass region. Since large parts of the search (based on an integrated luminosity of $189\,\mathrm{fb}^{-1}$) are still background-free, we can expect sensitivity improvements on the product of branching ratios by more than an order of magnitude. Concretely, we take the published expected sensitivities for the muon, pion and kaon final state and assume that future searches can improve sensitivity by a factor of 20 in each channel. Following Ref.~\cite{Batell:2009jf,winklerDecayDetectionLight2019}, we take $\text{BR}(B^+ \to K^+ \phi) = 0.44 \sin^2 \theta$ for $m_\phi \ll m_B - m_K$ as well as the total decay width and branching ratios of $\phi$ as shown in figure~\ref{fig:fit_result}. The resulting sensitivity projections are shown in figure~\ref{fig:sensitivities}.

We find that even though the expected sensitivity for the product of the branching ratios is worse in the kaon final state than in the muon final state, the larger branching ratio predicted by our hadronisation model means that this channel is nevertheless competitive and promises the best sensitivity for scalar masses close to 2 GeV. For larger scalar masses, the branching ratio into $K^+ K^-$ drops rapidly in favour of final states with higher meson multiplicities, such that the muon channel once again offers the best sensitivity. The pion channel is less favourable due to a lower sensitivity than the muon channel and a smaller branching ratio than the kaon channel.

\section{Conclusions}

In this work we have considered the partial decay widths and branching ratios of a CP-even scalar boson that mixes with the SM Higgs boson. We have focused on the mass range $2 \, \mathrm{GeV} \leq m_\phi \leq 2 m_\tau$, for which neither dispersion relations nor perturbative calculations can be directly applied.

As a first step, we have re-evaluated perturbative calculations of the inclusive decays $\phi \to gg$ and $\phi \to s\bar{s}$. We find that the effect of including finite masses for bottom and charm quark, which is crucial at leading order, gives only a small correction at NLO. We therefore use N$^4$LO results from the literature for the decay widths assuming one infinitely heavy and three massless quarks. This approach yields a significantly larger total decay width than previous estimates, resulting in a suppression of the leptonic branching ratios and the lifetime of the scalar.

To obtain branching ratios into specific hadronic final states, we have developed a hadronisation model based on the cluster model of \textsc{Herwig}. In contrast to the existing implementation, we assume that the two gluons produced in the scalar decay form a single cluster, which decays into a pair of mesons. To select the allowed final states, we check conservations laws such as parity and charge conjugation. These final states are assigned a weight according to the available phase space and selection rules for spin and isospin.

Our hadronisation model introduces two free parameters: the strange-quark weight and the suppression of final states with orbital angular momentum. These parameters can be fixed by matching the predictions of our model to those from dispersion relations at $m_\phi = 2 \, \mathrm{GeV}$. This procedure makes it possible to extrapolate the partial decay widths into pions and kaons to $m_\phi > 2 \, \mathrm{GeV}$. Moreover, the hadronisation model also predicts branching ratios for other states that may be of experimental interest, such as $\phi \to 2\pi^+ 2\pi^-$.

We find that for $m_\phi$ slightly above 2 GeV, both $\text{BR}(\phi \to \pi^+ \pi^-)$ and $\text{BR}(\phi \to K^+ K^-)$ are larger than $\text{BR}(\phi \to \mu^+ \mu^-)$, making these final states an attractive target for future searches. To illustrate this point, we have estimated the sensitivity of a future search for these final states at Belle II. We conclude that the kaon final state does indeed hold the potential to cover additional parameter space of a Higgs-mixed scalar not testable with the muon final state.

Our study should be considered as a proof-of-principle for the use of hadronisation models with conservation laws. There are many possible directions in which this work can be extended. For example, it could be interesting to allow for cluster fission, such that more than high meson multiplicities can be produced, or to include baryonic decay modes. Moreover, our model could be extended to include off-shell effects for mesons with a large width, as well as mixing of $\phi$ with CP-even scalar resonances. In fact, the latter effect may not only modify the branching ratios, but also the total width and hence the lifetime of the scalar.

Finally, our approach relies strongly on the results from dispersion relations, which in turn require input from experimental data. Improving these measurements could give tighter constraints on the partial decay widths below $2 \, \mathrm{GeV}$, which in turn would lead to reduced uncertainties on the free parameters of our hadronisation model. Combining these experimental and theoretical efforts, there is a real chance to make substantial progress on the long-standing problem of the decay modes of a GeV-scale CP-even scalar.

\section*{Acknowledgements}

We are very grateful to Jan Jerhot for comments on the manuscript and to Michael Spira for help with \textsc{higlu}. We furthermore thank  Patrick Ecker, Torben Ferber, Simon Knapen, Ulrich Nierste, Maksym Ovchynnikov, Michele Papucci, Peter Reimitz, Rhitaja Sengupta, Matthias Steinhauser,  Jaume Tarrús Castellà and Jure Zupan for discussions. FK acknowledges funding from the Deutsche Forschungsgemeinschaft
(DFG) through Grant No. 396021762 -- TRR 257.

\appendix

\section{Selection Rules} \label{app:selection_rules}

In this appendix, we examine the conserved quantities and the resulting selection rules for hadronic decays of a scalar particle $\phi$ that is even under both parity and charge conjugation. We assume that these decay are mediated by the strong interaction, where parity, charge conjugation and isospin are conserved. As in our hadronisation model, we consider only decays into meson pairs, neglecting both decays into baryons and decays into higher-multiplicity final states.

\subsection{Parity} \label{sec:parity}

If a particle state has orbital angular momentum (OAM) $l$, then the spatial wave function of this system has parity $(-1)^l$. As a result, a meson with intrinsic OAM $l$ has parity
\begin{equation}
 P=(-1)^{l+1},  \label{eq:Pmeson} 
\end{equation}
where the additional factor of $(-1)$ comes from the quark having opposite parity to the antiquark.
For a system of two particles with OAM $l$ and individual intrinsic parities $P_1$ and $P_2$ the total parity is
\begin{equation}
    P=P_1P_2(-1)^l \; . \label{eq:Ptwoparticles}
\end{equation}
If $\phi$ decays into two scalar mesons, conservation of angular momentum implies $l = 0$, such that both mesons must have the same parity. For all decays to mesons of higher spin, the allowed intrinsic parities of both mesons depend on their relative OAM (see below).

\subsection{Charge conjugation} \label{sec:Csymmetry}

Since $\phi$ is even under charge conjugation, every charged\footnote{The term "charged" refers here to electrical charge as well as flavour. All particles that are not their own antiparticles are considered to be charged.} particle produced in the decay of $\phi$ must be accompanied by its antiparticle, in order for the two-particle state to be a $\widehat{C}$-eigenstate. This requirement strongly restricts the possible decays and excludes all decays into two different charged mesons such as $\pi^+\rho^-$. Decays to two different neutral mesons such as $\eta\eta'$ are still possible, in which case $C$ is the product of the $\widehat{C}$-eigenvalues of the individual mesons.

The eigenvalue under charge conjugation for a particle-antiparticle state is known from their total spin $s$ and their relative angular momentum $l$: $C=(-1)^{l+s}$~\cite{palIntroductoryCourseParticle2014}. This equation automatically gives $C = 1$, since the total spin and OAM have to add to a total angular momentum of zero.

\subsection{Angular momentum} \label{sec:angularmom}

All decays have to respect angular momentum conservation, meaning that the total angular momentum of the final state must be zero. 
This requirement restricts the possible spin configurations of the final states and therefore determines the weight $W_v$ representing the spin multiplicity. We denote the spins of the two mesons as $j_1$ and $j_2$, their combined spin as $J$ and the OAM as $l$. The following cases are possible:
\begin{itemize}
\item $j_1=j_2=j$ and $P_1=P_2$:
    This combination can always couple to zero angular momentum, even without relative OAM. 
    The Clebsch-Gordon coefficients for coupling two angular momenta to zero total angular momentum are given by \cite{biedenharnAngularMomentumQuantum1981}
    \begin{equation}
    \braket{j,m_1;j,m_2}{0,0} = \delta_{m_1,-m_2} (-1)^{j-m_1} \frac{1}{\sqrt{2j+1}}.
    \end{equation}
    In other words, if there is no relative OAM, the state of zero total angular momentum is a superposition $2j+1$ states, each with the same amplitude. We therefore set $W_v = 2j + 1$ in this case. If non-zero OAM is allowed, all possible combinations of the meson spins, i.e. of $m_1$ and $m_2$, are allowed. The result is therefore a superposition of all $(2j+1)^2$ possible states.

    \item $j_1=j_2=j$ and $P_1=-P_2$: In this case $l$ must be odd to conserve parity, which also requires $J$ to be odd. Under these conditions, the Clebsch-Gordon coefficients for $m_1 = m_2$ vanish, such that only $2j(2j+1)$ allowed final states remain. We neglect this relatively minor effect and take the spin multiplicity to be $(2j+1)^2$ also if the two mesons have opposite parity. 
    \item $j_1=0$ and $j_2\neq0$: The OAM has to be $l=j_2$ to allow the total spin and OAM to couple to zero. Hence, if $j_2$ is odd, both mesons must have opposite intrinsic parities to satisfy parity conservation and if $j_2$ is even they must have the same parities. This excludes for example decays to $\pi^0\rho^0$, because they have the same parity. 
    The multiplicity of the state is always $(2j_2+1)$.
\item $j_1\neq j_2$ and both non-zero: No general conclusions can be drawn. In this case, we take the multiplicity as $(2 j_1 + 1)(2 j_2 + 1)$, which also includes the case that $j_1 = 0$ from above.
\end{itemize}
To summarize, if we require no OAM, only final states with $j_1 = j_2$ and $P_1 = P_2$ are allowed with weight $W_v = (2 j_1 +1)$. If we impose no restriction on OAM, the weight is (approximately) $W_v = (2j_1+1)(2j_2+1)$. Introducing the parameter $a_v$ to suppress OAM then leads to the expression for $W_v$ given in the main text.

\subsection{Isospin} \label{sec:isospin}

Isospin is conserved approximately by the strong interaction, if the difference between up and down quark masses is neglected. Since the scalar $\phi$ has $I=0$, the isospins of the two particles resulting from the $\phi$ decay have to couple to zero, which is only possible if they are the same. This excludes for example decays to the combination $\pi^0$ ($I=1$) and $\eta$ ($I=0$), which would be otherwise allowed by $P$ and $C$ conservation. 

In contrast to the case of angular momentum, the $I_3$ component is always fixed for any meson by the contained quarks. As a result, decays into mesons with higher isospin are not enhanced by a multiplicity factor, but are in fact suppressed compared to the decay to isoscalars. If both particles have isospin $I$, the probability that they couple to an isospin singlet is given by $1/(2I + 1)$. For example, decays into pairs of kaons ($I = \tfrac{1}{2}$) are suppressed by a factor of $\tfrac{1}{2}$, while decays into pairs of pions\footnote{The three pions are identified with the isospin states $\pi^+=\ket{1,1}, \pi^-=\ket{1,-1}, \pi^0=\ket{1,0}$} or pairs of $\rho$-mesons ($I = 1$) are suppressed by factor of $\tfrac{1}{3}$.

\subsection{G-parity} \label{sec:gparity}

The G-parity transformation describes a rotation around the $I_2$ axis in isospin space followed by a charge conjugation:
\begin{equation}
    \hat{G}=\hat{C}\exp{i\pi \hat{I}_2}.
\end{equation}
Since charge conjugation and isospin are conserved individually in strong interactions, $G$ is conserved as well.
All mesons that contain only up and down quarks are eigenstates of this transformation, and all mesons in the same isospin multiplet carry the same eigenvalue. For a meson multiplet of isospin $I$ the eigenvalue is given by $G=(-1)^IC$, where $C$ is the eigenvalue of the neutral member of the multiplet under charge conjugation. Since $\phi$ has $I=0$ and $C=1$, it can be assigned an eigenvalue of $G=1$. G-parity conservation then excludes all decays into mesons with $G=-1$. Since pions are G-parity odd, while $\rho$ mesons are G-parity even, this conservation law forbids decays into $\pi\rho$ or any odd number of pions.

\normalem

\end{document}